\documentstyle[aps,preprint]{revtex}

\begin{document}

\newcommand{\be}{\begin{equation}}
\newcommand{\ee}{\end{equation}}
\newcommand{\bea}{\begin{eqnarray}}
\newcommand{\eea}{\end{eqnarray}}
\newcommand{\bc}{\begin{center}}
\newcommand{\ec}{\end{center}}
\newcommand{\eq}[1]{(\ref{#1})}
\newcommand{\tw}{t_{\rm w}}
\newcommand{\eqn}[2]{\begin{equation}\label{#1} #2 \end{equation} }
\newcommand{\avg}[1]{\left\langle #1 \right\rangle }
\newcommand{\xe}{x_{\rm e}}
\newcommand{\Ce}{C_{\rm e}}
\newcommand{\peq}{p^{\rm eq}}

\draft

\title{Off-Equilibrium Dynamics of 1+1 dimensional
Directed Polymer\\ in Random Media}

\author{Hajime Yoshino}
\address{Institute of Physics,
University of Tsukuba, Tsukuba, Japan\\
{\rm email: yoshino@cm.ph.tsukuba.ac.jp}
}

\maketitle

\begin{abstract}
The relaxational dynamics of 1+1 dimensional directed
polymer in random potential is
studied by Monte Carlo simulations.
A series of temperature quench experiments is performed changing
waiting times $\tw$.
Clear crossover from quasi-equilibrium behavior ($t \ll \tw$)
to off-equilibrium behavior ($t \gg \tw$) appears in the dynamical
overlap function whose scaling properties are very similar to
those found in the 3 dimensional spin-glass model.
In the $t \gg \tw$ part, the fluctuation dissipation theorem of the 1st kind
which relates the response function to the tilt field with the conjugate
correlation function, is found broken. These aging effects are brought
about by the very slow growth of quasi-equilibrium domain driven by successive
{\it loop-excitations} of various sizes, which form complex network
structures.
\end{abstract}


\newpage
\section{Introduction}

Directed polymer in random media (DPRM)\cite{HZ95} is one of the simplest
statistical mechanical models in which quenched randomness
plays non-trivial roles as in spin-glasses.
It is an effective model of an elastic string in random environment,
such as a vortex line in oxide cuprate
superconductor which penetrates the stacked ${\rm CuO}_{2}$ layers
with point defects scattered randomly over the layers.
In transverse dimensions $d$ less than two, there is no
{\it free phase} but a {\it pinned phase},
i.e. the polymer is mostly pinned around the
ground state and cannot move freely at any finite temperature.
However there are
anomalously large thermal fluctuation due to the thermal hoppings between
the excited states which are nearly degenerate
with the ground state but located far away.
They are analogous to the droplet excitations
in spin-glass phase\cite{FH88} and bring about non-trivial
effects in the pinned phase\cite{FH91,HF94,M90}.
One naturally expects that
they have dramatic effects also on the dynamics.
Actually, it has been argued in the theory of the transport
problem in the vortex glass phase that they are responsible for the
nonlinear current-voltage response\cite{vortexglass}.

Our interest in the present study
is the slow relaxation to equilibrium due to such anomalous
thermal hoppings. We expect that they bring about
aging effects, which are to some extent
similar to those found in the spin-glass phase.
In order to clarify this possibility,
we performed temperature
quench experiments by a simple heatbath Monte Carlo dynamics.
The procedure mimic the so called IRM experiments in spin-glasses\cite{LSNB}
but the system is perturbed by a small {\it tilt} field which
drives one end of the polymer instead of magnetic field in spin-glasses
which drives the whole spins .

We have found clear evidences of the aging effects.
One is the systematic waiting time
dependence on the dynamical overlap function. The crossover
behavior from quasi-equilibrium to off-equilibrium behavior and
its scaling properties are, interestingly enough,
very similar to that found in the spin-autocorrelation function
of the 3 dimensional spin-glass model\cite{R93}.
The 2nd is the break down of the fluctuation dissipation theorem (FDT)
associated with the response to the tilt field.
Again the correlation function which is conjugate with the
response function shows clear waiting time dependence.
On the other hand, in contrast with the case of spin-glasses, the response
function itself does not show waiting time dependence.
The FDT which relates the two breaks down at $t \gg \tw$,
which coincide with
the crossover behavior in the dynamical overlap function.

In our analysis we also utilize the transfer matrix method for
the following two purposes.
Firstly, we check the consistency between the
static limit of the data of the dynamical quantities and the static
expectation values.
Secondly we study the
relation between the complex free energy landscape
and the relaxational dynamics. Actually we could visually
monitor the thermal jumps between the excited states.
The structure of the  web of the excited states is very complex
which consists of numerous loop-like structures
of various sizes (see Fig.~3 below).
Since this model is rather simpler than spin-glass models, we believe
that understanding
the dynamics of this model would give valuable insight into
the glassy dynamics of more complicated systems.

This paper is organized as the following. In section 2
we describe our model and introduce the Monte Carlo dynamics.
In section 3, the elementary process
of the dynamics is studied combining with the analysis of the
free energy landscape by the transfer matrix method.
In section 4, the procedure of the
temperature quench experiment is described. The results are
given in section 5.
In section 6 we conclude this paper with some
phenomenological arguments.

\section{The Model}

We study a lattice version of
1+1 dimensional directed polymer in random media (DPRM).
The Hamiltonian is
\eqn{H}{{\cal H}_{\mu}[x] =  \sum_{z=1}^{N} \left\{
g|x(z)-x(z+1)|+\mu(x(z),z)
\right\}
-hx(N)}
where $x(z)$ $(z=1\cdots N)$ represents the configuration of
the polymer whose length is $N$.
As shown in Fig.~1, the polymer is situated on a square lattice
and {\it directed} along the $z$-axis so that
overhangs are excluded. We impose the so called
restricted solid on solid (RSOS) condition,
i.e. the steps $|x(z)-x(z-1)|$ are only allowed to take
integer values $-1, 0$ and $1$.
We fix one end $x(0)$ at $x=0$ while we allow the other end $x(N)$
to move freely. The 1st term in the
Hamiltonian represents the elastic energy.
The 2nd term represents the random potential
which takes random numbers distributed uniformly over
$-\sigma\leq\mu\leq\sigma$
independently on every lattice site on the square lattice.
The last term is the tilt field which drives the free end
and {\it tilts} the polymer.

We model the relaxational dynamics of the present model by an
ordinary heatbath Monte Carlo dynamics
as in the following way. In one micro Monte Carlo step (micro-MCS),
one segment, say the segment at $z=z_{0}$ is tried to move. The new
position is chosen among the possible choices allowed by
the RSOS condition
with the appropriate transition probabilities
so as to satisfy the detailed balance condition.
As shown in Fig.~2, this micro-MCS
can be classified into three cases depending on the positions of
the neighboring segments at $z_{0}-1$ and $z_{0}+1$, namely
case  a), b) and c) which corresponds to
$|x(z_{0}-1)-x(z_{0}+1)|=0$
,$1$ and $2$ respectively.
The entire configuration is refreshed by vectorized sublattice flippings
in one Monte Calro step (MCS) (=N micro-MCS).
One sublattice consists of  the even numbered segments
$z=2,4,6\ldots $ and the other sublattice consists of the
odd numbered segments $z=1,3,5\ldots$.

\section{Elementary processes of the dynamics}

In this section we consider the elementary processes of
the relaxational dynamics in the present model.
In any quenched random systems, it is generally essential to
understand the structure
of the free energy landscape in order to study
the slow relaxational dynamics, which is unfortunately
very hard to do on spin-glass models. However,
thanks to the simplicity of the model,
we can try such an attempt on the present
model in the following way.
We consider the spatial variation of $P_{\mu}(z,x)$:
the probability to find
the polymer going through a particular lattice point $(z,x)$
on a certain sample of random potential $\mu$.
It is defined as
\eqn{land}{P_{\mu}(z,x)= Z_{\mu}^{-1}\sum_{{\rm configuration}}
\delta(x(z)-x)\delta(x(0)-0) \exp(-\beta{\cal H}[x,\mu] )}
where the sum is taken over the all possible configurations
(or paths) $\{x(1),\ldots, x(N)\}$ and
$Z_{\mu}$ is the normalization factor
defined so that $\sum_{x=-z}^{z} P_{\mu}(z,x)=1$.
It is straightforward to calculate this probability
by transfer matrix method \cite{HZ95}.
In Fig.~3, an example of the spatial variation of $P_{\mu}(z,x)$
of a system of $N=100$, $T=0.3$, $g=0.01$ and $\sigma=0.5$
on a certain sample of random potential $\mu$  is displayed
by a density plot;
the intensity increases as the color becomes brighter.
We recognize that most of the probability density is confined in
the white {\it tubes} which are understood as the thermally active
excited states.
In the dynamics, they presumably serve as {\it traps} \cite{B92},
which tend to trap the polymer for long times.
On the other hand there are
black {\it voids} between the tubes which presumably
serve as free energy barriers. It is quite remarkable that
the tubes form a very complicated network which consists of
various sizes of loop-like structures.
It is qualitatively consistent with the prediction by
Hwa and Fisher\cite{HF94} who have predicted a broad distribution
of the size of the loops.
As the temperature increases, one can observe that
the diameter of the tubes grows
and the structure itself changes on large scales
as predicted in\cite{FH91}.

{}From the above observation, we speculate that
the dynamics of the polymer roughly consists of the two kinds of
elementary processes.
One is the fast fluctuation within each tubes (traps).
The other is the thermally activated jumps between such tubes,
which is analogous to the droplet excitations in spin-glasses.
Actually these two different processes can be visually
monitored rather easily as the following.
In Fig.~4 we plot the time sequence of the segment $x(80)$ on the same
system shown in Fig.~3.
The data are averaged over every interval of $10^3$ (MCS)
trying to mask small scale fluctuations.
Apart from the remanent small scale fluctuations, which is
understood as the fast fluctuations within each traps (tubes),
the coarsegrained process is apparently understood as jumps between
the four traps which correspond to
the four major peaks of $P(80,x)$ (white tubes in Fig.~3 at $z=80$).

The thermal jumps between the traps must go over the free energy barriers
which lie between them (black void in Fig.~3).
The scaling property of the relaxation times of
the thermal jumps between the tubes has already been
studied in the context of the aforementioned
transport problem in the vortex glass phase\cite{vortexglass}.
For the convenience of the later discussions, we
summarize it here.
For example, suppose that there is
a loop-like structure of tubes as schematically shown in Fig.~5.
Let its transverse size $\Delta$.
Then the free energy barrier $F_{\rm B}(\Delta)$ associated with the
thermal excitation over the loop, which we hereafter call
{\it loop-excitation} following\cite{vortexglass},  scales as
\eqn{scaleFb}{F_{\rm B}(\Delta) \simeq a(T) \Delta^{\alpha},}
where $a(T)$ is some temperature dependent constant.
Note that near the free end of the polymer, the tubes does not
form complete loop structures but rather {\it U}-shaped structures
in which one side of the ends of the tubes are left open.
But we expect that
such thermal excitations scales in the same way as the {\it complete}
loop excitations.
The value of the exponent $\alpha$ is expected to be
$1/2$ in 1+1 DPRM\cite{HF94}.
Then the relaxation time of the loop-excitation scales as
\eqn{looptau}{\tau(\Delta) \sim \exp(F_{\rm B}(\Delta)/T)
\simeq \exp(\frac{a(T)\Delta^{\alpha}}{T}).}
The broad distribution of the size of the loops\cite{HF94}
implies broad distribution of relaxation times.

\section{Procedure of the Temperature Quench Experiment}

The numerical experiment is done in the following way (see also
the illustration in Fig.~6).
At first, a random initial configuration is prepared.
Then the system is let to evolve
for $\tw$ (MCS) by the heatbath Monte Carlo dynamics of
a certain temperature $T$.
This means that the polymer is forced to approach the equilibrium
state of temperature $T$ being attached to a heat bath of
temperature $T$.
After this $\tw$ (MCS) of {\it aging}, the polymer's configuration is stored
and copied to a replica system.
Then we let the two systems, say replica A and B, continue the
relaxational dynamics using
the same random numbers (same thermal noise of the heat bath)
but appling small tilt field $h$ to system B.

We measure the following
dynamical overlap function as a probe to
the relaxational dynamics of the unperturbed system (replica A).
It is defined as
\eqn{q}{q(\tw+t,\tw) \equiv \overline{\avg{\frac{1}{N}\sum_{z=1}^{N}
 \delta(x^{(A)}(z,\tw+t), x^{(A)}(z,\tw))}},}
where $\delta(x, y)$ is the Kroneker's delta
and $x^{(A)}(z,t)$ ($z=1\ldots N$)
represents the configurations of the polymer  at $t$ (MCS).
Here the bracket $\left\langle\ldots  \right\rangle$
denotes the avarage over different sequences of the random
numbers and the over bar $\overline{\cdots}$
the average over different realizations of
the random potentials. It is defined in an analogy with the
spin-autocorrelation function of the spin-glass models
and conveniently measure the 'closeness' between the configurations
at two different times $\tw+t$ and $\tw$.
In the static limit, it is expected to converge to the
expectation value of the overlap $q_{\rm rep}$ between two replica systems
which has been introduced by M\'{e}zard\cite{M90},
\eqn{staticq}{q_{\rm rep}=q_{\rm static}\equiv\lim_
{t \rightarrow \infty}\lim_{\tw \rightarrow \infty}
q(\tw+t,\tw).}

We measure the dynamical response to the tilt field at $t$ (MCS)
after $\tw$ (MCS) of aging
as the distance between the temporal positions of the free end $x(N)$
of the two replicas A and B, which
we denote as $x_{e}^{(A)}(\tw+t)$ and $x_{e}^{(B)}(\tw+t)$,
\eqn{resp}{\overline{\delta{\xe(t;\tw)}}
\equiv \overline{\avg{\xe^{(B)}(\tw+t)-\xe^{(A)}(\tw+t)}}.}
Note that $\left\langle\ldots  \right\rangle$ encloses both
$x_{e}^{(A)}(t)$ and $x_{e}^{(B)}(t)$ because we use the same random
numbers in replica A and B in our procedure.
This definition is slightly different from the
conventional one $\overline{\avg{\xe^{(B)}(\tw+t)}}
-\overline{\avg{\xe^{(A)}(\tw+t)}}$.
However our present choice is more efficient in practice because
it provides essentially the same result but with less noise.

In equilibrium, the response function \eq{resp} is related
with the correlation function
\eqn{C1}{\Ce(t_{1},t_{2})=\overline{\avg{\xe^{(A)}(t_{1})\xe^{(A)}(t_{2})}}}
as
\eqn{balance}{\overline{\delta{\xe(t;\tw)}} = h \int_{\tw}^{\tw+t}
R_{\rm e}(t+\tw,\acute{t}) d\acute{t}\\
= \beta h \left[ \Ce(\tw+t,\tw+t)-\Ce(\tw+t,\tw)
\right],}
by the fluctuation disspation theorem (FDT) of the 1st kind,
\eqn{fdt}{R_{\rm e}(t_{1},t_{2})\equiv \overline{
\delta{\xe(t_{1}-t_{2};t_{2})}}/\delta h(t_{2})\\
=\beta\partial \Ce(t_{1},t_{2})/\partial t_{2}}
provided that $h$ is small enough (linear response regime).
Note that in the static limit
$\tw \rightarrow \infty$ and $t \rightarrow \infty$
of \eq{balance}, we
get the ordinary static FDT,
\eqn{staticresp}{\overline{\avg{\xe}_{h}}=
\lim_{t \rightarrow \infty}\lim_{\tw \rightarrow \infty}\overline{
\delta{\xe(\tw;t)}}=\beta h \chi}
where the susceptibility $\chi$ is given by
\eqn{chi}{\chi=
\overline{\avg{\xe^{2}}-\avg{\xe}^{2}}.}

We performed a series of temperature quench experiments on the system
with various sizes $N=20 \ldots 300$,
and temperatures at $T=0.1 \ldots 0.5$. The parameter $g$ and $\sigma$ are
fixed at $g=0.01$ and $\sigma=0.5$.
The average over $10^4\ldots 10^2$ samples
of random potential were done depending on the system size.
In order to ensure that $h$ is small enough so that the
linear response condition is satisfied, we calculated
the disorder averaged static susceptibility $\chi$ (see \eq{chi}) and
$\overline{\avg{\xe}_{h}}$ by the transfer matrix method\cite{M90}
and checked that \eq{staticresp} is well satisfied.

\section{Result}
\subsection{Crossover Behavior of Dynamical Overlap Function}

We first present the results of
the dynamical overlap function defined in \eq{q}
on the unperturbed system (replica A). The following analysis is
done on the data of $N=300$ at $0.1 < T < 0.5$.
The observation was done in the time window ($0< t <10^{4}$ (MCS))
for $\tw=2,2^{2},\ldots ,2^{10}$ (MCS).
The data at $T=0.1$ and $0.3$ are plotted by
double logarithmic plot in Fig.~7.
The remarkable feature is that
the data curves show strong waiting time $\tw$ dependence
and each of them
drops off rapidly at around $t \sim \tw$.
It is presumably understood as a
manifestation of the crossover
from the quasi-equilibrium behavior
($t\ll\tw$) to the off-equilibrium behavior ($t\gg\tw$), i.e. aging effect.
Actually we see later that it coincides with the FDT breaking which
also occurs at around $t \sim \tw$.

The quasi-equilibrium regime ($t \ll \tw$) begins with
an initial fast decay and
crossovers to a much slower decay,
which is only visible on the data curves of sufficiently large $\tw$.
Unfortunately we cannot determine
the asymptotic functional form ($\tw \gg 1$)
of the latter slow decay at this moment.
However the scaling analysis which we discuss later suggests
an algebraic law (see \eq{quasieq} below) with very small
exponent $x(T)$ which varies with the temperature as shown in Fig.~10.
On the other hand, the off-equilibrium part is well fitted by an algebraic law
\eqn{offeq}{q(\tw,\tw+t) \sim t^{-\lambda(T)}\ \ \ \ \ \ (t \gg \tw),}
where the exponent $\lambda(T)$ also depends on temperature as
shown in Fig.~10.
It is interesting to note that these characteristics of the
crossover behavior are very similar
to those originally found by Rieger\cite{R93}
in the spin-autocorrelation function of
the 3 dimensional Edwards-Anderson spin-glass model
in the spin-glass phase.

We now analize the scaling properties of the
dynamical overlap function.
We try the scaling form
\eqn{scaleq}{q(\tw,\tw+t)\simeq C(\tw,T)\tilde{q}_{T}(t/\tw).}
As an example, the scaling plot of the data of $N=300$, $T=0.17$ is shown in
Fig.~8.
The parameter $C(\tw,T)$ was chosen so as to scale the
off-equilibrium part ($t/\tw \gg 1$) as good as possible.
It turns out that the initial fast decay part, which we mentioned before,
does not fit on the master curve of the $t/\tw$ scaling.
In practice, we could obtain the scaling plot shown in Fig.~8
discarding the data of $t \le 10$ (MCS).
At higher temperatures we have to discard more data to obtain good
master curves.

As expected, the scaling function
behaves as $\tilde{q}_{T}(y) \sim y^{-\lambda(T)}$ at $y \gg 1$
with $\lambda(T)$ which we found in \eq{offeq}.
On the other hand the constant $C(\tw,T)$ turns out to be well
fitted by an algebraic law of $\tw$,
\eqn{constC}{C(\tw,T) \sim \tw^{x(T)},}
as shown in Fig.~9.
The exponent $x(T)$ obtained by this fitting varies with
temperature as presented in Fig.~10.
At $T < 0.17$, $x(T)$ becomes too small to be determined
precisely.

Using \eq{constC}, we
can rewrite \eq{scaleq} as
\eqn{scaleq_R}{q(\tw,\tw+t)\simeq t^{-x(T)}\phi_{T}(t/\tw)}
where $\phi_{T}(y)$ is related with $\tilde{q}_{T}(y)$ in \eq{scaleq} by
$\phi_{T}(y)\equiv y^{x(T)}\tilde{q}_{T}(y)$. Note that this scaling form is
also identical to that originally found in 3 dimensional spin-glass model by
Rieger\cite{R93}. The new scaling form \eq{scaleq_R} implies that the
quasi-equilibrium part ($t/\tw \ll 1$)
behaves as
\eqn{quasieq}{q(\tw,t+\tw) \sim t^{-x(T)}\ \ \ \ \ \
(\tw \gg 1\ \ , \ \   t/\tw \ll 1).}
However as can be seen in the master curve in Fig.~8, the crossover
takes place rather gradually and the left branch of the master curve
($t/\tw \ll 1$)
seem to decay faster than the expected algebraic law $t^{-x(T)}$ with
$x(T)$ obtained from \eq{constC}.
This discrepancy implies that we have not yet attained the asymptotic
scaling behavior of the $t/\tw \ll 1$ part.
Unfortunately we could not accomplish it with our available computational
power.

Lastly we discuss the finite size effects. As far as the system
size $N$ is finite, the size of the largest
loop-excitation and thus the associated maximum relaxation
time $\tau_{\rm max}(N)$ available in the system must be finite
(see \eq{looptau}).
Thus we expect that system is fully equilibrated
if we take $\tw$ larger than a certain equilibration time
$t_{\rm eq}(N)$ which may be comparable with the $\tau_{\rm max}(N)$.
The finite size effect appears in the dynamical overlap function
as the following.
In Fig.~11 the data on the system of $N=20$ and $T=0.5$ is shown as
a typical example.
It can be seen that
the waiting time $\tw$ dependence
on the dynamical overlap function saturates at some finite $\tw$ as
$\tw$ is increased:
the curves do not show any further $\tw$ dependence when $\tw$ exceeds
some characteristic time $t_{\rm eq}(N)$. In the example shown
in Fig.~11, it is larger than
$2048$ but less than $16384$ (MCS).
We have checked that
$t_{\rm eq}(N)$ increases as we increase the system size $N$.
It is also recognized that the dynamical overlap
function saturates to the static limit $q_{\rm static}$ (see \eq{staticq})
when $t \gg t_{\rm eq}$.
The value of $q_{\rm static}$ is nearly equal to $q_{\rm rep}$
obtained by transfer matrix method which uses two replicas\cite{M90}.
Though $q_{\rm rep}$ is non-zero in finite systems,
further calculations by the transfer matrix method on larger systems
show that $q_{\rm rep}$
decrease as the system size $N$ increases and
it vanish in the thermodynamic limit $N \rightarrow \infty$
(see\cite{M90}).

\subsection{Breakdown of FDT}

We next present the results of
the response to the tilt field defined in \eq{resp}.
The curves in Fig.~12 shows the l.h.s
and r.h.s. of the FDT relation \eq{balance} at different waiting times
$\tw=40$, $200$ and $1000$ (MCS)
on the system of $N=20$ at $T=0.2$
with the tilt field $h=0.1$ by semi-logarithmic plot.
The quench experiment was done over
$10^4$ different samples of random potential in order to take
sufficient average over the disorder.
The curves of the l.h.s. of \eq{balance} (correlation)
strongly depends on the waiting time $\tw$.
One the other hand, the curves of  the r.h.s of \eq{balance} (response)
of corresponding $\tw$  do not seem to have
significant $\tw$ dependence and almost seem to marge with each other
within the numerical accuracy.
Comparing with the data of larger system sizes, we checked that
the data are not spoiled by finite size effects within this
observation time window ($0 < t < 10^4$ (MCS)). However it is difficult
to get smooth data on larger systems  because they are not
self-averaging quantities and  sample to sample
fluctuations are larger on larger systems.

As is clear from the figure,
the FDT relation is satisfied in the regime
$t \ll \tw$ but  broken in the
regime $t \gg \tw$.
It is again a clear evidence of the crossover from quasi-equilibrium
to off-equilibrium dynamics, i.e. aging effect.
Similar FDT breaking was previously found in the spin-glass phase of the
3 dimensional spin-glass models by Monte Calro simulations\cite{AMS,FR95}.
However there is an important difference: in the spin-glass
models both the response and correlation functions strongly
depends on $\tw$ while in the present model the response dose not
seem to have $\tw$ dependence. This is understood as the following.
As the tilt field pulls the free end of the polymer, it moves by
successive local thermal excitations around it,
while the rest of the system is
indifferent of the existence of the tilt field.
We expect that the free energy barrier of such
thermal excitations scales as \eq{scaleFb}.
Thus the response will scale the same way as
the {\it typical} transverse size of a
loop-excitation which becomes active after waiting $t$,
\eqn{scaleresp}{\overline{\delta{\xe(t;\tw)}} \sim \log(t)^{1/\alpha}.}
In order to clarify this scaling form, we performed longer simulation
up to $10^{5}$ (MCS).
In Fig.~13 we show the double
logarithmic plot of $\overline{\delta{\xe(t;0)}}$ v.s. $\log(t)$
on the system of $N=30$, $T=0.20$ and $N=60$ at $T=0.30$ with
$h=0.03$.
It can be seen that the behavior is consistent with \eq{scaleresp}
with $\alpha=1/2$ and thus supports the above argument.
The data on the system of different system sizes and
at different temperatures show similar behavior except that data of
smaller systems show saturations within the observation time window.

Lastly we consider the extremely high temperature case in finite
systems.
As we mentioned before  the diameter of the tubes
grow as the temperature increases\cite{HF94}.
Then if the system is too small, a single tube swallows up
the whole system and the thermal jumps between the tubes disappear.
In Fig.~14 is shown an example of free energy landscape for such case,
the density plot
of $P_{\mu}(z,x)$ of a system of $N=20$ at $T=1.0$ on a certain random
potential. Roughly only a single big tube can be recognized.
In such an extreme case, we found no aging effects.
The FDT  relation  is fulfilled in the whole $t$ range
up to the static limit as shown in Fig.~15.
At around $10^3$ (MCS), they saturate to the static limit,
$\overline{\avg{\xe}_{h}}$
or $h/T\chi_{e}$ (see \eq{staticresp})
calculated by the  transfer matrix method.
In addition, the dynamical overlap function does not show any $\tw$ dependence.
These facts suggest that the
thermal jumps between the tubes, i.e. loop-excitations are
the essential processes to bring about the observed aging effects.

\section{Conclusion}

The relaxational dynamics of 1+1 dimensional DPRM was investigated by
the Monte Carlo simulations which mimiced the IRM experiment in spin-glasses.
The aging effect appears as the systematic waiting time
dependence on both the dynamical overlap function and the FDT breaking.
The dynamical overlap function shows clear crossover
from slow quasi-equilibrium decay ($t \ll \tw$)
to fast off-equilibrium decay ($t \gg \tw$).
The latter is well fitted by ab algebraic law whose exponent depends
on temperature. The $t/\tw$ scaling scheme works well and
it suggests algebraic decay also in the
quasi-equilibrium part. However we could not confirm the latter
directly with our computational power.
On the other hand, the response to the tilt field grows as
$\log(t)^{2}$ \eq{scaleresp} independently of the waiting time,
which supports
the assumed scaling properties of loop-excitations
(see \eq{scaleFb} and \eq{looptau}).
Concerning the fact that it does not show $\tw$ dependence,
this model appears rather trivial compared with, for example,
spin-glasses in which there are dramatic effects also
in the response function so that one can actually observe
the aging effects in experiments by measuring
the magnetization \cite{LSNB,RVHO}.
None the less, other aspects of the aging effect
are very similar to that found in the spin-glass phase of
3 dimensional spin-glass models\cite{R93,AMS,FR95}.
Hence we believe that this model provides a good testing ground
for various theoretical ideas of the slow dynamics of quenched random systems.

The important elementary processes to bring about the observed aging effect
are the loop-excitations, i.e. thermal hoppings between the traps (tubes)
in the free energy landscape. They were actually demonstrated
by direct monitoring of the time sequences.
Thus equilibration process is understood as the slow growth of
quasi-equilibrium domain driven by successive loop-excitations.
Note that this growth mechanism of the quasi-equilibrium domain
is essentially the same as proposed in the spin-glass models
in which droplet excitations drive the growth of the domain\cite{FH88}.
{}From \eq{looptau}, we expect that its transverse size $R(t)$ grows as
\eqn{domain}{R(t) \sim \log(t)^{1/\alpha}.}
However the scaling properties of the dynamical overlap function
are not trivial.
Though they turned out to be quite similar to those found in
the spin-glass model\cite{R93}, they cannot be explained by
simply borrowing the assumptions
of the above mentioned droplet theory\cite{FH88}.
In the droplet theory, it is assumed that
\eqn{qR}{q(t,0) \sim R(t)^{-\lambda},}
where $q(t_{1},t_{2})$ is now the spin-autocorrelation function
and $\lambda$ is some unknown exponent\cite{FH88}.
(Here $\lambda$ is the notation of \cite{FH88} and different
from $\lambda(T)$ defined in \eq{offeq}.)
But we don't have any reason to believe that such a relation
exists in the present model, and actually our result \eq{offeq}
cannot be consistent with \eq{domain} if we assume \eq{qR}.
In the study of spin-glass model\cite{R93},
it is argued that \eq{offeq} can be consistent with \eq{qR}
if one assumes that the free energy barrier grows logarithmically
with its transverse size $\Delta$ rather than algebraically
so that $R(t)$ grows algebraically with
a certain temperature dependent exponent.
If it is the also case in the present model, the response function \eq{resp}
should be fitted by such algebraic law.
However our data of the response function show systematic curvature
in double logarithmic plots and do not prefer such algebraic law but
the logarithmic law \eq{scaleresp}.
Thus we conclude that the simple relation \eq{qR}
does not exist in the present model. The scaling properties of
the dynamical overlap function require other ways to explain them.

The important feature we recognize in Fig.~3 is that the
tubes are grouped together to form some very complicated network.
We already know how the relaxation time associated with each single
loop-excitation scales with its transverse size (see \eq{looptau}).
However there is an obvious but important rule: a loop-excitation
can {\it flip} only when some part of the temporal configuration of
the polymer actually stay in either side of the loop, in other words
the {\it empty} loops can not {\it flip}. It should also be noted
that the polymer is {\it not} allowed to be broken into pieces
during the dynamical process.
Thus not the entire loop-excitations built in the system
can become {\it active} simultaneously but only those which
are associated with the sequence of the loops which contain the
actual temporal configuration of the polymer.
{}From the above considerations, it is clear that
the information of the connectivity of the loops is an important
ingredient to describe the dynamics of the polymer.

Let us now consider the dynamical overlap function \eq{q}.
It can be roughly interpreted as the {\it probability}
that the segments of the polymer return to
the original traps in which they initially stay after jumping around
other traps in the network.  Here we need
the detailed description of the dynamics mentioned above
since the probability is the sum of the probabilities associated with
all of the possible paths which make such return trips.
Each of these paths consist of different sequences of
loop-excitations which vary in  sizes and thus have different relaxation times.
It is unlikely that there is a single {\it typical} mode whose probability
dominate the total probability.
On the other hand, the response to the tilt field does not
depend directly on such details. The response function scales just as the
{\it expectation value} of the transverse size of
the local loop-excitation around the free end which can become active
after waiting $t$. It is possible that such expectation value is dominated
by the contribution from the {\it typical} mode whose relaxation time
is comparable with $t$. We think this is the reason why we could obtain
\eq{scaleresp} using only the scaling property of single loop-excitation.

Although the network looks apparently quite complicated,
a most simple characterization of it will be the following.
The network structure consists of loops of various
sizes which are hierarchically nested so that larger ones
enclose smaller ones inside.
(A similar picture was proposed before by Villain
concerning the organization of {\it droplets}
in spin-glass phase, see\cite{V86}).
{}From this simple picture, the dynamics can be regarded
as one of the hierarchical diffusion processes which
have been intensively studied in rather abstract contexts,
for example the {\it ultradiffusion} models\cite{udiff}.
It is interesting to note that in the latter models
the autocorrelation function (return probability)
shows asymptotically algebraic decay with some temperature dependent
exponent, which is also the case in the dynamical overlap function
in the present model.
Furthermore, it has already been shown by Sibani and Hoffman that
hierarchical diffusion processes can show aging effects\cite{SH}.
Thus it will be fruitful to construct a phenomenological theory
of the aging effect of the present model
from this point of view.
A study in this direction will be reported elsewhere\cite{Y}.

\vspace*{5mm}
\noindent
Acknowledgement: The author would like to thank sincerely
Prof. H. Takayama for valuable discussions and critical reading
of the manuscript. He would also like to thank Dr. H. Reiger,
Prof. K. Nemoto, K. Hukusima and T. Komori for stimulating discussions.
This work was supported by Grand-in-Aid for Scientific Research
from the Ministry of Education, Science and Culture, Japan.
The author was supported by Fellowships of the Japan Society
for the Promotion of Science for Japanese Junior Scientists.

\newpage
\noindent
{\bf \large FIGURE CAPTIONS}

\vspace*{3mm}\noindent
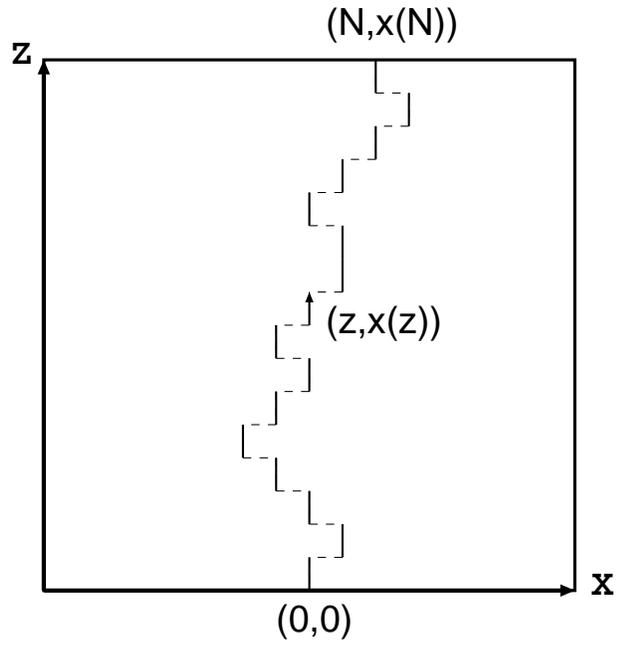
Fig.~1: A representation of a configuration of 1+1 dim. RSOS DPRM.

\vspace*{3mm}\noindent
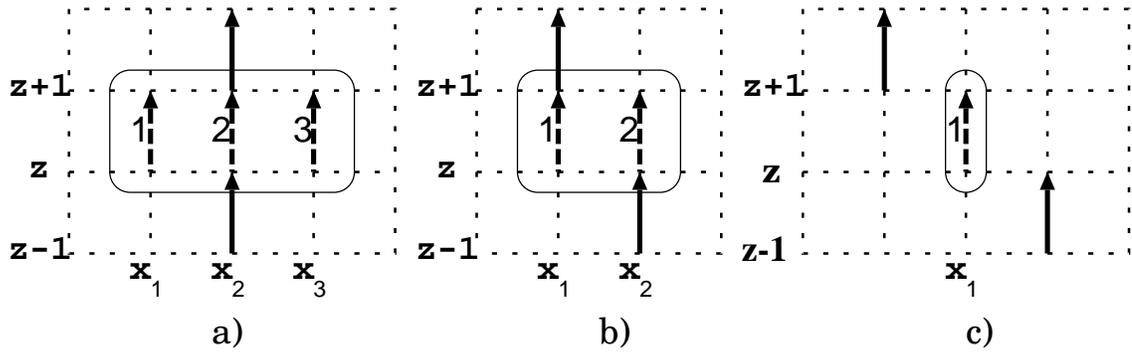
Fig.~2: The three different cases a), b) and c) of the local
configuration around a segment (see text for the definition).
The possible new states $1,2$ and $3$ of the segment are enclosed
by boxes.

\vspace*{3mm}\noindent
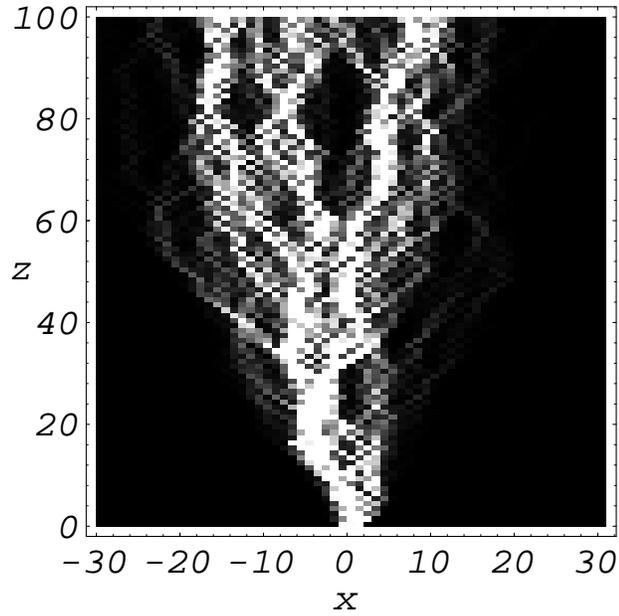
Fig.~3: The density plot of
$P_{\mu}(z,x)$ (see text for definition)
on a certain sample of
random potential $\mu$
($N=100$, $T=0.3$, $g=0.01$ and $\sigma=0.5$).

\vspace*{3mm}\noindent
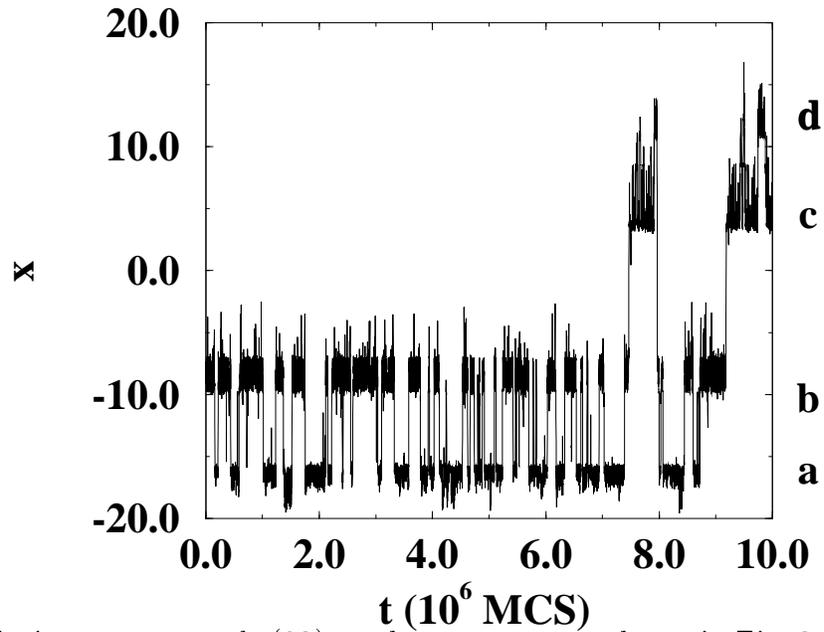
Fig.~4: A time sequence of $x(80)$
on the same system shown in Fig.~3.
The positions of the four traps a, b, c and d are indicated.

\vspace*{3mm}\noindent
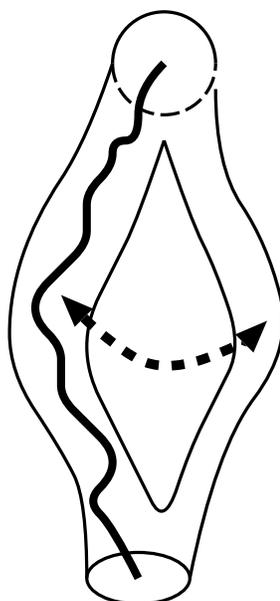
Fig.~5: A schematic picture of a loop-excitation.
The polymer (bold line) inside the left tube
jumps over the void to the right tube and vice versa.

\vspace*{3mm}\noindent
Fig.~6: The procedure of the temperature quench experiment.

\vspace*{3mm}\noindent
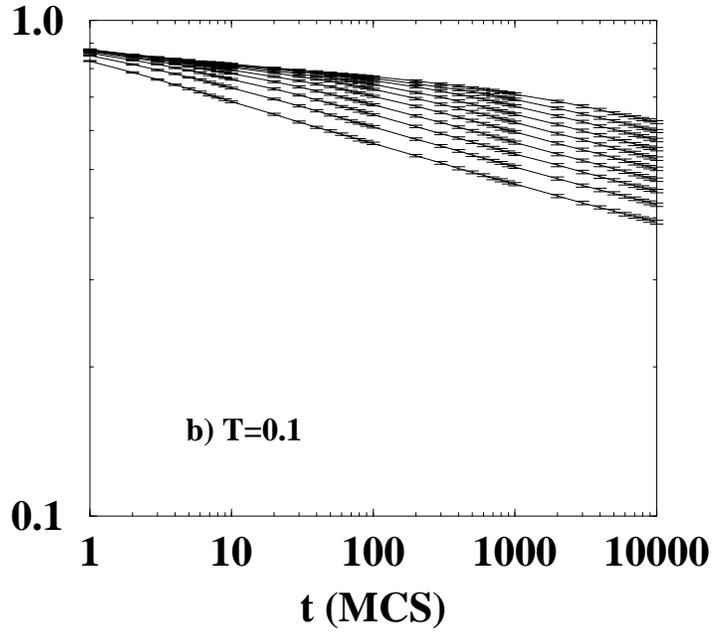
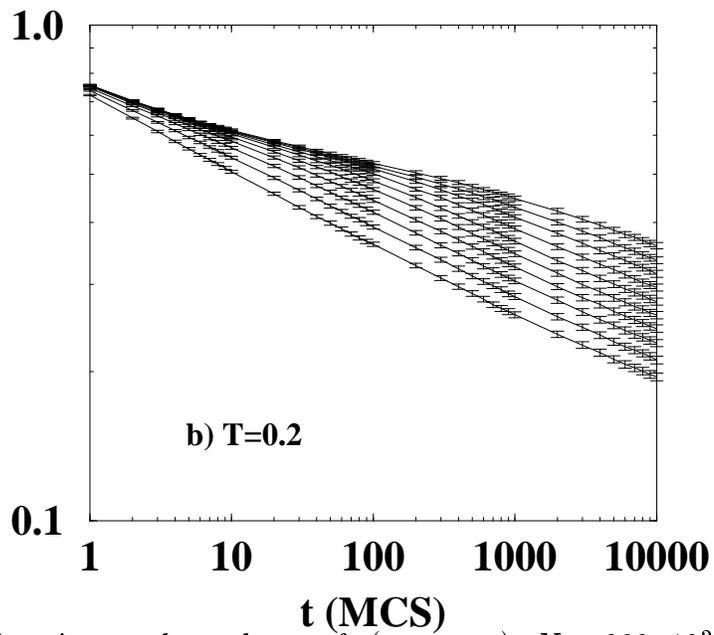
Fig.~7: The waiting time $\tw$ dependence of $q(\tw,\tw+t)$:
$N=300$, $10^2$ samples at a) $T=0.1$ and  b) $T=0.2$.
The waiting time increases as $\tw=2,4,8,...1024$ (MCS) from the bottom
to the top curve.

\vspace*{3mm}\noindent
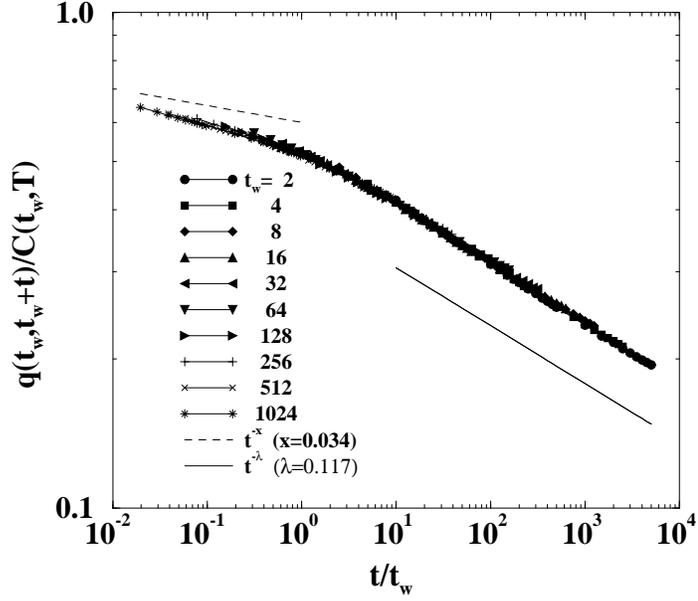
Fig.~8: A scaling plot of the dynamical overlap function
on the system of $N=300$ and $T=0.17$  by the scheme \eq{scaleq}.
The scale of the vertical axis is arbitrary. The doted line represents
the power law  $t^{-x}$ using $x$ obtained by the fit \eq{constC}.
The solid line represents the power law  $t^{-\lambda}$ with
$\lambda$ obtained by the direct fit of the $t/t_{\rm w} \gg 1$ part.

\vspace*{3mm}\noindent
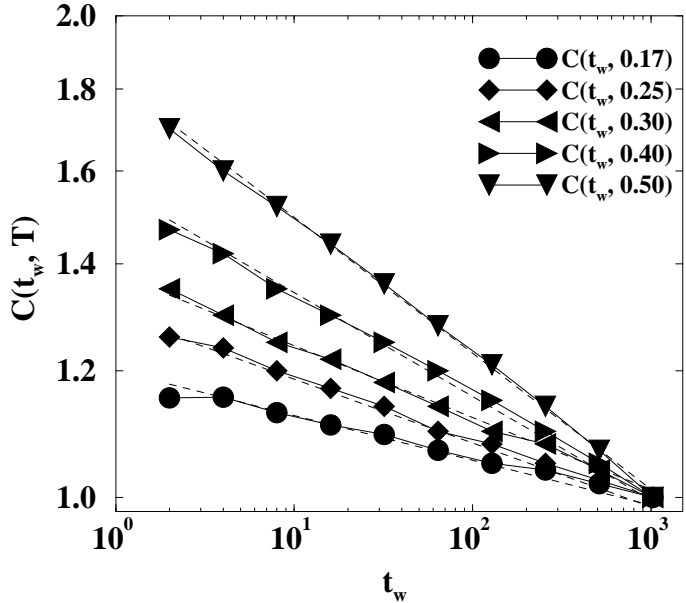
Fig.~ 9: Double logarhithmic plot of $C(\tw,T)$ v.s. $\tw$ at
$T=0.17, 0.25, 0.30, 0.40$ and $0.50$.
The scale of the vertical axis is arbitrary.
The dashed lines are the fit by the algebraic law \eq{constC}
with $x(T)$ presented in Fig.~10.

\vspace*{3mm}\noindent
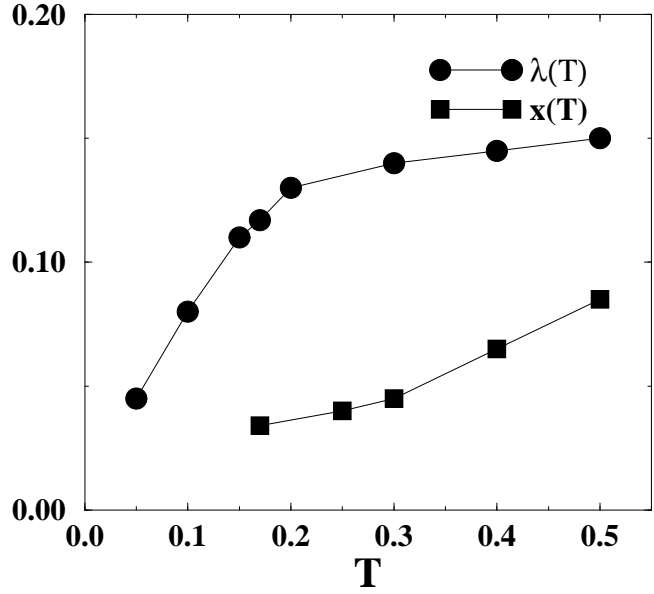
Fig.~10: Temperature dependence of the exponents
: $\lambda(T)$ and $x(T)$. They are obtained from data of
($N=300$, $10^2$ samples).

\vspace*{3mm}\noindent
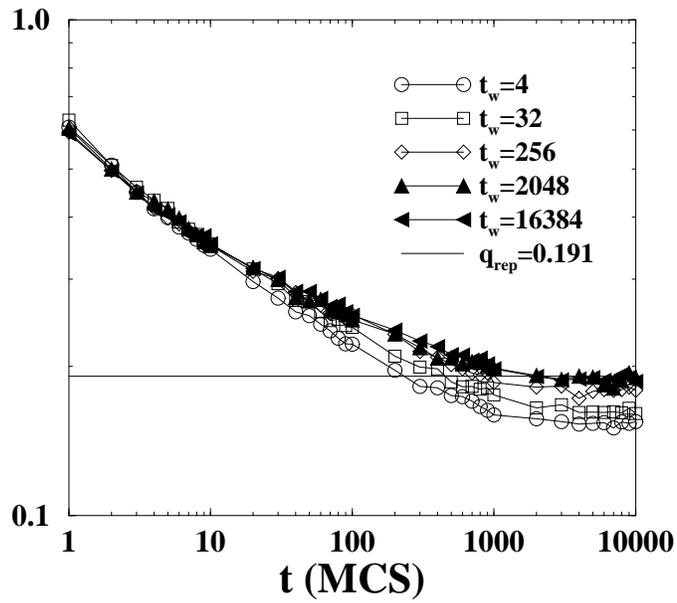
Fig.~11: Saturation of aging
effect due to the finite size effect
($N=20$, $T=0.5$, $500$ samples). $q_{\rm rep}$ was
calculated by the transfer matrix method.

\vspace*{3mm}\noindent
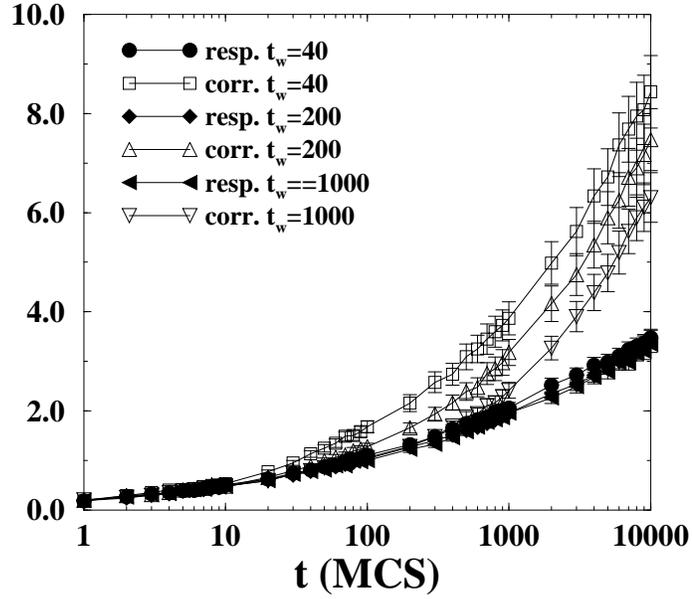
Fig.~12: FDT relation
($N=20$ $T=0.2$  $10^4$ samples).
The top three curves are
the l.h.s. of \eq{balance} (correlation)
of different waiting times $\tw=40,200$ and $1000$ (MCS).
The three curves below
are the r.h.s of \eq{balance} (response)
of corresponding $\tw$.

\vspace*{3mm}\noindent
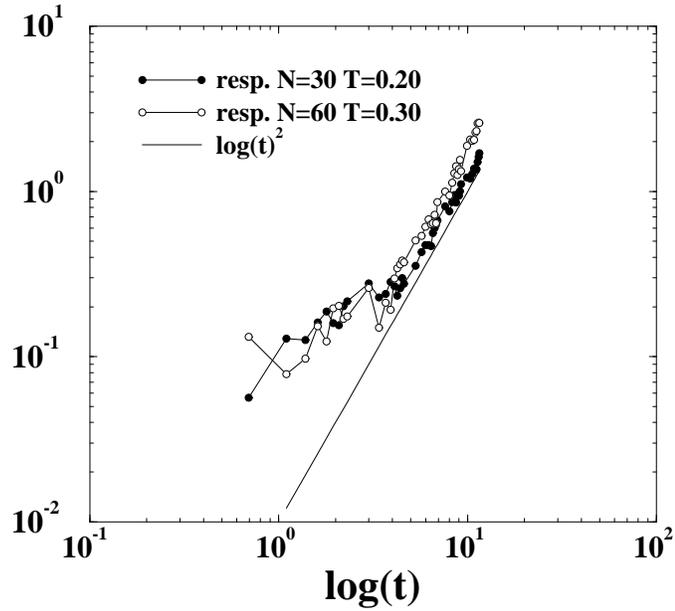
Fig.~13: Functional form of the response function
($N=30$, $T=0.20$, $10^4$ samples
and $N=60$, $T=0.40$, $5\times 10^3$ samples with $h=0.03$).
The solid line represents $\log(t)^{2}$ law \eq{scaleresp}.

\vspace*{3mm}\noindent
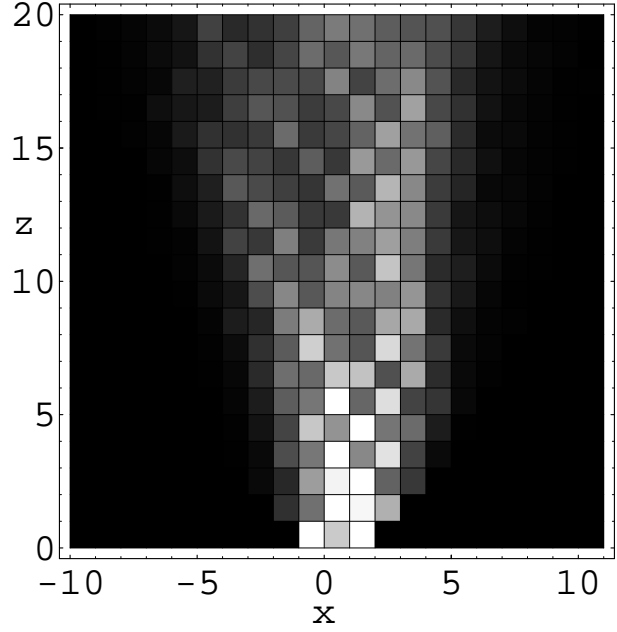
Fig.~14: The density plot of
$P(z,x)$ on a certain sample of small system
at a high temperature.
($N=20$, $T=1.00$)

\vspace*{3mm}\noindent
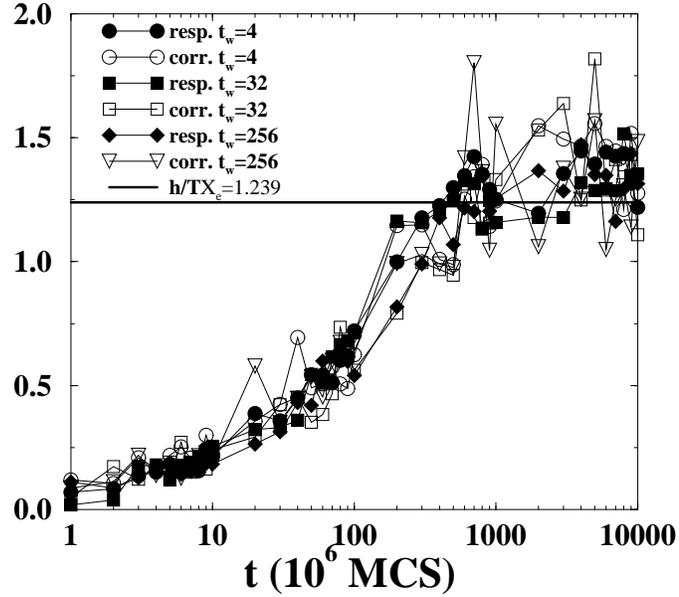
Fig.~15: The FDT relation \eq{balance} of a small system
at a heigh temperature ($N=20$, $T=1.00$, $h=0.1$, $500$ samples)
for $\tw=4,32,256$ (MCS).
The horizontal line is $h/T \chi_{e}$ where $\chi_{e}$
(see \eq{chi}) is
obtained by transfer matrix method.

\end{document}